\documentclass[conference]{IEEEtran}
\IEEEoverridecommandlockouts

\usepackage{cite}
\usepackage{amsmath,amssymb,amsfonts}
\usepackage{amsthm}
\usepackage{algorithm}
\usepackage{algorithmic}
\usepackage{textcomp}
\usepackage{xcolor}
\usepackage{booktabs}
\usepackage{tabularx}
\usepackage{array}
\usepackage{url}
\usepackage{stfloats}
\usepackage{balance}
\usepackage{flafter}

\usepackage{soul}
\usepackage{multirow}
\usepackage{graphicx} 

\newcommand{\Comment}[1]{/* {#1} */}

\newcommand{\cmitoprule}[1]{\cmidrule[\heavyrulewidth]{#1}}

\newcolumntype{Y}{>{\raggedright\arraybackslash}X}
\newcommand{\calX}{\mathcal{X}}

\newcommand{\Var}{\mathrm{Var}}
\newcommand{\KL}{\mathrm{KL}}

\newcommand{\sign}{\mathrm{sign}}

\newtheorem{definition}{Definition}
\newtheorem{proposition}{Proposition}
\newtheorem{lemma}{Lemma}
\newtheorem{theorem}{Theorem}
\newtheorem{corollary}{Corollary}

\def\BibTeX{{\rm B\kern-.05em{\sc i\kern-.025em b}\kern-.08em
    T\kern-.1667em\lower.7ex\hbox{E}\kern-.125emX}}

\begin{document}

\title{A Bernstein-Type Variance-Adaptive PMW Mechanism for DP Query Release}

\author{\IEEEauthorblockN{Mohamed Aniss Benstaali}
\IEEEauthorblockA{\textit{\'Ecole Nationale Sup\'erieure d'Informatique}\\
Algiers, Algeria}
\and
\IEEEauthorblockN{Habib Louafi}
\IEEEauthorblockA{\textit{T\'ELUQ University}\\
Montreal, Canada}
\and
\IEEEauthorblockN{Naima Bessah}
\IEEEauthorblockA{\textit{\'Ecole Nationale Sup\'erieure d'Informatique}\\
Algiers, Algeria}
}

\maketitle

\begin{abstract}
Private Multiplicative Weights (PMW) is a central mechanism for the differentially private release of adaptive linear statistics over a finite universe. Its classical analysis uses a fixed multiplicative-weights learning rate and does not exploit the local variance of the current hypothesis. This paper presents PMW-Bernstein, a variance-adaptive refinement in which the update step is determined by the relationship between the privately estimated query error and the variance under the current hypothesis. The mechanism retains the standard PMW privacy architecture based on Sparse Vector screening, Gaussian numerical measurements, sequential composition, and post-processing. Through a potential-based analysis, we prove a one-step progress bound that yields a global weighted energy bound, recovers the classical $O(1/\alpha^2)$ update regime in general, and establishes an $O(1/\alpha)$ update regime under a Bernstein-type condition on updated rounds. Controlled simulations over synthetic finite-universe distributions and query workloads show that the adaptive rule reduces the number of updates in the low-variance regime, remains competitive in the high-variance regime, and succeeds within the prescribed update budget in the heterogeneous regime where the fixed-step baseline fails.
\end{abstract}


\section{Introduction}

Organizations, such as healthcare systems and financial institutions, increasingly rely on repeated statistical analyses of sensitive data~\cite{dwork2014algorithmic}. A hospital, for instance, may need to report disease prevalence, treatment outcomes, and subgroup statistics from a patient database. In interactive settings, analysts may choose subsequent queries after inspecting earlier answers; therefore, the system must support an adaptive sequence of queries rather than a single predefined one~\cite{hardt2010multiplicative}.

However, an uncontrolled number of queries can be exploited by attackers to infer sensitive data by accumulating the answers to previous queries. Therefore, the number of queries is usually limited (controlled) through a \textit{privacy budget}. The privacy budget bounds the cumulative privacy loss incurred by answering queries, thereby preventing attackers from extracting excessive data through repeated or adaptive interactions with the database. Moreover, exact answers may reveal sensitive information about individual records. Adding calibrated noise to query answers preserves privacy but consumes the limited privacy budget through composition, which may eventually reduce the utility of the released statistics. This highlights the fundamental privacy--utility trade-off of the problem at hand~\cite{dwork2014algorithmic}.

Therefore, differential privacy (DP) provides a framework for managing this trade-off by bounding cumulative privacy loss through a privacy budget while limiting the influence of any single database record on the distribution of released query answers~\cite{dwork2006calibrating,dwork2014algorithmic}.

In this context, several techniques have been proposed for answering queries under DP, including fundamental mechanisms such as the Laplace and Gaussian mechanisms~\cite{dwork2006calibrating,dwork2014algorithmic} and the Exponential Mechanism~\cite{mcsherry2007mechanism}. Moreover, advanced query-answering approaches, such as Private Multiplicative Weights (PMW)~\cite{hardt2010multiplicative}, the Matrix Mechanism~\cite{li2010matrix}, and Fourier-based contingency-table release~\cite{barak2007privacy}, have been proposed to address complex scenarios, including large query workloads, correlated queries, and interactive or adaptive query answering.

PMW maintains a public hypothesis distribution over a finite universe and uses this hypothesis to produce approximate query answers. A new private measurement is obtained only when the hypothesis is found to be inaccurate, and the privacy budget is updated accordingly~\cite{hardt2010multiplicative}. PMW combines two key ideas: the Sparse Vector Technique, which limits the number of costly accesses to the private database, and multiplicative updates, which progressively reduce the discrepancy between the hypothesis and the data distribution represented by the private database.

The classical PMW analysis uses an update magnitude determined by global accuracy parameters rather than by the local variance of the current query under the current hypothesis. Consequently, the correction does not explicitly distinguish between queries whose values are highly dispersed under the hypothesis and queries that are locally concentrated but still exhibit a large model discrepancy.

To address this gap, we propose a Bernstein-type variance-adaptive refinement of the PMW update rule. The privacy primitives and their budget allocation remain standard; the new component is the multiplicative-update magnitude, which depends on the relationship between the privately estimated query error and the local variance under the current hypothesis. The resulting rule applies a stronger correction when the hypothesis is locally concentrated but inaccurate and a more moderate correction when the query has larger local variance.

The main contributions are as follows:
\begin{enumerate}
\item We propose a PMW-Bernstein refinement mechanism that retains standard private screening and noisy numerical measurements while replacing the fixed update magnitude by a variance-adaptive multiplicative step.
\item We establish the differential privacy guarantee of the proposed mechanism by accounting for the privacy costs of the screening and numerical-measurement stages under standard composition, resulting in a bounded cumulative privacy loss.
\item We derive a weighted energy bound that characterizes the cumulative progress of the mechanism and show that, under the Bernstein-type variance-error condition, the number of required updates scales as $O(1/\alpha)$, where $\alpha$ is the target accuracy.
\item We demonstrate the feasibility and performance of the proposed mechanism through controlled simulations over a finite data universe using synthetic data distributions and query workloads.
\end{enumerate}

The rest of the paper is organized as follows. Section~\ref{sec:related-work} reviews related work on PMW, DP, and adaptive update mechanisms. Section~\ref{sec:preliminaries} introduces the preliminaries and notation used throughout the paper, followed by the problem statement in Section~\ref{sec:problem-statement}. Section~\ref{sec:proposed-solution} presents the proposed PMW-Bernstein framework. Section~\ref{sec:privacy-analysis} establishes its DP guarantee. Section~\ref{sec:potential-analysis} develops the potential-based analysis of the adaptive update and derives the corresponding update-complexity bounds. Section~\ref{sec:key-components-derivation} provides the derivation of the key components of PMW-Bernstein. Section~\ref{sec:privacy-utility-tradeoff} discusses the resulting privacy-utility trade-off. Section~\ref{sec:experimental-setup} describes the experimental methodology, while Section~\ref{sec:experimental-results} presents the simulation results. Section~\ref{sec:discussion} discusses the experimental findings. Finally, Section~\ref{sec:conclusion} concludes the paper.

\section{Related Work}
\label{sec:related-work}

This section reviews the main approaches to repeated differentially private query release. The discussion covers direct perturbation and privacy accounting, sparse screening methods, and learning-based mechanisms that maintain a reusable public approximation of the private data. Their respective design properties clarify the need for an update rule that accounts for the local geometry of each detected error.

\subsection{Direct perturbation and privacy accounting}

The Laplace and Gaussian mechanisms perturb sensitivity-bounded statistics directly and provide simple, modular privacy guarantees \cite{dwork2006calibrating,dwork2014algorithmic}. When applied independently to a long query sequence, however, each release contributes to the cumulative privacy loss. Advanced composition, concentrated DP, and R\'enyi DP provide sharper accounting for repeated releases \cite{dwork2010boosting,bun2016concentrated,mironov2017renyi}, but they do not reduce the number of private measurements required by the interaction. This motivates mechanisms that can reuse previously acquired information to answer subsequent queries.

\subsection{Sparse private screening}

The Sparse Vector Technique (SVT), also known through AboveThreshold variants, screens a long stream of sensitivity-bounded tests while limiting the privacy cost associated with threshold-crossing events~\cite{dwork2014algorithmic,lyu2017understanding}. This makes SVT well suited to identifying the rounds on which a public approximation is inaccurate. Its output is a private threshold decision, so a complete release mechanism must additionally specify how a selected query is measured and how the public state is corrected after that measurement.

\subsection{Multiplicative weights and private data release}

Learning-based data-release mechanisms construct a synthetic or hypothesis distribution that can answer many queries without further access to the private database. Blum et al. established a learning-theoretic approach to noninteractive release \cite{blum2008learning}, while the median mechanism supports interactive query answering by maintaining a set of databases consistent with previous measurements \cite{roth2010interactive}. Hardt and Rothblum introduced the Private Multiplicative Weights (PMW) mechanism, which combines sparse detection with multiplicative updates of a public hypothesis \cite{hardt2010multiplicative}. The Multiplicative Weights Exponential Mechanism (MWEM) later combined private query selection and measurement with multiplicative-weights updates in a practical iterative procedure \cite{hardt2012simple}; its update is driven by the measured discrepancy between the selected query and the current hypothesis. More recent iterative methods introduce other forms of adaptivity. Private Entropy Projection (PEP) reuses previously measured queries through an entropy-based projection step \cite{liu2021iterative}, while the Adaptive Iterative Mechanism (AIM) adaptively selects useful marginal measurements for private synthetic-data generation \cite{mckenna2022aim}. These mechanisms show that adaptivity can enter query selection, measurement reuse, or state reconstruction. The specific question studied here is different: how to choose the multiplicative-update magnitude from the current measured discrepancy and the query variance under the current hypothesis, together with a variance-sensitive potential guarantee for that choice.

\subsection{Synthesis of Existing Approaches}

Prior work addresses different aspects of adaptive private query release through different mechanisms. Direct perturbation provides basic privacy primitives, while privacy budget consumption scales with every measurement. Sparse screening helps identify rounds in which additional private measurements are needed, but it does not determine the actual update. Hypothesis-based and synthetic-data approaches introduce adaptivity through query selection, measurement reuse, or state reconstruction. These approaches, however, do not directly address the choice of the multiplicative-update magnitude as a function of the current query discrepancy and local query variance.

In contrast, this work develops a variance-sensitive update rule and establishes a corresponding Bernstein-type one-step potential bound. PMW-Bernstein retains the standard screening and measurement components of the PMW mechanism while introducing variance-sensitive update dynamics and a formal analysis of these dynamics. Table~\ref{tab:related-work-comparison} presents a comparison of different approaches proposed in the literature that are related to the problem considered in this work.

\begin{table*}[t]
\caption{Comparison of representative approaches to differentially private query release and the proposed PMW-Bernstein mechanism.}
\label{tab:related-work-comparison}
\centering
\begin{tabular}{l c c c c c c}
\toprule
\multirow{2}{*}{\textbf{Approach}} & \textbf{Private} & \textbf{Reusable} & \textbf{Adaptive} & \textbf{Measurement} & \textbf{Adaptive} & \textbf{Variance-aware} \\
& \textbf{screening} & \textbf{state} & \textbf{query selection} & \textbf{reuse} & \textbf{update magnitude} & \textbf{update} \\
\midrule
Laplace/Gaussian mechanisms & $\times$ & $\times$ & $\times$ & $\times$ & $\times$ & $\times$ \\
Sparse Vector / AboveThreshold & $\checkmark$ & $\times$ & $\times$ & $\times$ & $\times$ & $\times$ \\
Median Mechanism & $\times$ & $\checkmark$ & $\times$ & $\times$ & $\times$ & $\times$ \\
PMW & $\checkmark$ & $\checkmark$ & $\times$ & $\times$ & $\times$ & $\times$ \\
MWEM & $\times$ & $\checkmark$ & $\checkmark$ & $\times$ & $\checkmark$ & $\times$ \\
PEP & $\times$ & $\checkmark$ & $\checkmark$ & $\checkmark$ & $\checkmark$ & $\times$ \\
AIM & $\times$ & $\checkmark$ & $\checkmark$ & $\checkmark$ & $\times$ & $\times$ \\
\textbf{Proposed: PMW-Bernstein} & $\checkmark$ & $\checkmark$ & $\times$ & $\times$ & $\checkmark$ & $\checkmark$ \\
\bottomrule
\end{tabular}
\end{table*}

\section{Preliminaries and Notations}

\label{sec:preliminaries}

In this section, we introduce the notation and definitions used throughout
the paper, including empirical distributions, substitution adjacency,
differential privacy, and query sensitivity.

\begin{definition}[Empirical distribution]
Let $\mathcal{X}$ be a finite universe, and let
$D=(x_1,\ldots,x_n)\in\mathcal{X}^n$ be a database of fixed size $n$.
The empirical distribution of $D$ is defined by
$$
p_D(x)=\frac{1}{n}\sum_{i=1}^n\mathbf{1}\{x_i=x\},
\qquad x\in\mathcal{X}.
$$
Thus, $p_D\in\Delta(\mathcal{X})$, where $\Delta(\mathcal{X})$ denotes
the probability simplex over $\mathcal{X}$, defined as
$$
\Delta(\mathcal{X})
=
\left\{
u\in\mathbb{R}^{|\mathcal{X}|}:
u(x)\geq 0,\;
\sum_{x\in\mathcal{X}}u(x)=1
\right\}.
$$
\end{definition}


\begin{definition}[Substitution adjacency]
Let $\mathcal{X}$ be a finite universe, and let
$D=(x_1,\ldots,x_n)$ and $D'=(x'_1,\ldots,x'_n)$ be two databases
in $\mathcal{X}^n$ of fixed size $n$. The databases $D$ and $D'$ are
said to be substitution adjacent, denoted by $D\sim D'$, if there
exists a unique $i\in\{1,\ldots,n\}$ such that
$$
x_i\neq x'_i
\qquad\text{and}\qquad
x_j=x'_j\quad\forall j\neq i.
$$
In other words, $D'$ is obtained from $D$ by replacing exactly one
record with another record.
\end{definition}


\begin{definition}[Differential privacy]
A randomized mechanism $\mathcal{M}$ is $(\varepsilon,\delta)$-differentially
private if, for every pair of substitution-adjacent databases $D\sim D'$ and
every measurable set $S$ of possible outputs,
$$
\Pr[\mathcal{M}(D)\in S]
\leq
e^{\varepsilon}\Pr[\mathcal{M}(D')\in S]+\delta,
$$
where $\Pr[X]$ denotes the probability of the event $X$, $\varepsilon\geq 0$ is the privacy parameter that controls the multiplicative factor $e^{\varepsilon}$, and $\delta\in[0,1]$ is the additive slack that allows the privacy guarantee to hold up to a small failure probability.
\end{definition}


\begin{definition}[Normalized query]
A normalized query is a function $q:\mathcal{X}\to[0,1]$. Its answer on a database $D=(x_1,\ldots,x_n)$ is defined as
$$
q(D)=\frac{1}{n}\sum_{i=1}^n q(x_i) =\sum_{x\in\mathcal{X}}p_D(x)q(x).
$$
\end{definition}


\begin{proposition}[Sensitivity]
\label{prop:sensitivity}

Under substitution adjacency, any normalized query $q:\mathcal{X}\to[0,1]$ has global sensitivity
$$
\Delta q = \sup_{D\sim D'}|q(D)-q(D')|\leq\frac{1}{n}.
$$
\end{proposition}

\begin{proof}
Let
$D=(x_1,\ldots,x_n)$ and $D'=(x'_1,\ldots,x'_n)$ be substitution-adjacent
databases. By definition, there exists a unique $i\in\{1,\ldots,n\}$ such that $x_i\neq x'_i$ and $x_j=x'_j$ for all $j\neq i$. Therefore,
$$
\begin{aligned}
|q(D)-q(D')|
&=
\left|
\frac{1}{n}\sum_{j=1}^{n}q(x_j)
-
\frac{1}{n}\sum_{j=1}^{n}q(x'_j)
\right|\\
&=
\frac{1}{n}|q(x_i)-q(x'_i)|\\
&\leq\frac{1}{n},
\end{aligned}
$$
where the last inequality follows from $q:\mathcal{X}\to[0,1]$.
Taking the supremum over all substitution-adjacent databases gives
$$
\Delta q = \sup_{D\sim D'}|q(D)-q(D')|\leq\frac{1}{n}.
$$
\end{proof}

For a statistic \(q_t:\calX\to[0,1]\), let \(h_t\in\Delta(\calX)\) be the current hypothesis, let \(\mu_t=\langle q_t,h_t\rangle\) be the prediction, let \(a_t\) be the true empirical value, let \(e_t=|\mu_t-a_t|\), and let \(v_t=\Var_{h_t}(q_t)\) be the local variance under the current hypothesis. When a private measurement gives an estimated error \(\widehat e_t\), the adaptive learning rate is set to
\begin{equation}
    \eta_t=\log\left(1+\frac{\widehat e_t}{v_t}\right),
    \label{eq:intro_eta}
\end{equation}
with the convention \(\eta_t=0\) if \(v_t=0\). The quantity \(\eta_t\) controls the magnitude of the multiplicative correction. The ratio \(\widehat e_t/v_t\) produces a stronger update when the hypothesis is concentrated but inaccurate and a more moderate update when the query has high local variance, while the logarithm controls the growth of the step. Section~VII further shows that this choice optimizes the one-step lower bound used in the potential analysis.

\begin{definition}[Sparse Vector Technique]
\label{def:SVT}

Let $(g_t)_{t=1}^k$ be a sequence of adaptively chosen queries with sensitivity $\Delta_g$, and let $T$ be a predefined threshold. The Sparse Vector Technique (SVT) privately compares each query $g_t(D)$ with $T$ and identifies at most $C$ queries whose noisy responses exceed the threshold. For the SVT variant considered in this paper, the threshold and test noises are sampled as
$$
\xi_j\sim\operatorname{Lap}\left(\frac{2C\Delta_g}{\varepsilon_{sv}}\right),
\qquad
\nu_t\sim\operatorname{Lap}\left(\frac{4C\Delta_g}{\varepsilon_{sv}}\right).
$$
\end{definition}

\section{Problem Statement}
\label{sec:problem-statement}

In interactive applications, such as medical data analysis, a data curator
holds a private database $D$ of $n$ records from a finite universe
$\mathcal{X}$ and answers a sequence of statistical queries submitted by an
analyst. The objective is to provide accurate answers across many
interactions while keeping the cumulative privacy loss within a prescribed
privacy budget. Technically, the problem consists of answering a long,
adaptive sequence of normalized statistical queries on $D$. Releasing every
answer independently with fresh noise can be inefficient because the privacy
loss accumulates across interactions.

The interaction with $D$ consists of $k$ rounds. At round
$t\in{1,\ldots,k}$, the analyst selects a normalized linear query,
possibly as a function of all previously released answers. The mechanism
then returns an answer to the query before proceeding to the next round.

Formally, let $q_t:\mathcal{X}\to[0,1]$ denote the query at round $t$.
Its empirical answer is
$$
a_t = q_t(D) = \frac{1}{n}\sum_{i=1}^{n}q_t(x_i) = \langle q_t,p_D\rangle,
$$

where $q_t(x)$ represents the bounded contribution of record $x$ to query
$q_t$, $p_D$ is the empirical distribution of $D$, and
$$
\langle q,u\rangle = \sum_{x\in\mathcal{X}}q(x)u(x)
$$

denotes the expectation of $q$ under a distribution $u\in\Delta(\mathcal{X})$.

The mechanism maintains a hypothesis distribution
$h_t\in\Delta(\mathcal{X})$, initialized from a prior
$r\in\Delta(\mathcal{X})$ that is public, chosen independently of the private database, and has full support, i.e.,
$r(x)>0$ for all $x\in\mathcal{X}$. At each round, $h_t$ represents the mechanism's current hypothesis distribution, obtained from the prior $r$ and updated using the information released in previous rounds. At round $t$, the corresponding
hypothesis answer is
$$
\mu_t=\langle q_t,h_t\rangle.
$$

We define the signed discrepancy, absolute error, and correction direction
as
$$
d_t=\mu_t-a_t, \qquad e_t=|d_t|, \qquad s_t=\operatorname{sign}(d_t),
$$

where $d_t$ measures the signed difference between the hypothesis answer
and the empirical answer, $e_t$ is the corresponding absolute discrepancy,
and $s_t$ indicates the direction in which the hypothesis should be
corrected.

The local variance of the query under the hypothesis distribution is
defined by
$$
\begin{aligned}
v_t
&=\operatorname{Var}_{h_t}(q_t)\\
&=\mathbb{E}_{x\sim h_t} \left[(q_t(x)-\mu_t)^2\right].
\end{aligned}
$$

Since $q_t(x)\in[0,1]$, the variance satisfies
$$
0\leq v_t\leq\frac14.
$$

The goal is therefore to design a mechanism that answers a sequence of adaptive
queries with target accuracy $\alpha$ while satisfying a prescribed
$(\varepsilon,\delta)$-differential privacy guarantee. Specifically, if the mechanism releases answers up to a stopping time $\ell\leq k$, we require, with high probability,
$$
|\widehat{a}_t-a_t|\leq\alpha \qquad \text{for all }t\in\{1,\ldots,\ell\}.
$$
A separate non-exhaustion condition is therefore needed when one wants to guarantee that all $k$ queries are processed before the update cutoff is reached.

Subject to these accuracy and privacy requirements, the mechanism aims to minimize the number of costly measurement rounds (i.e., rounds requiring fresh private measurements).

\section{Proposed PMW-Bernstein framework}
\label{sec:proposed-solution}

The proposed mechanism follows the PMW framework introduced for answering a sequence of adaptively chosen statistical queries under differential privacy. The main idea of PMW is to maintain a hypothesis distribution $h_t\in\Delta(\mathcal{X})$ that approximates the empirical distribution $p_D$ with respect to the queries received so far.

For a query $q_t$, the mechanism first computes the hypothesis answer $\mu_t=\langle q_t,h_t\rangle$. When this answer is sufficiently accurate, it can be released without performing a new private numerical measurement. When the discrepancy between the hypothesis answer and the database answer is large, a private measurement of the query is obtained, and the hypothesis is updated using a multiplicative-weights rule. This strategy allows the mechanism to answer a potentially long sequence of adaptive queries while limiting the number of rounds that require costly private numerical measurements.

The mechanism we propose builds on the PMW framework and incorporates a Bernstein-style variance-adaptive update. In particular, the update step uses the local query variance $v_t=\operatorname{Var}_{h_t}(q_t)$ to determine a variance-adaptive step size.

The resulting mechanism combines three components: (i) sparse private screening, (ii) private numerical measurement, and (iii) variance-adaptive multiplicative updates.

Let $\alpha\in(0,1]$ be the target accuracy, let $\varepsilon_{sv}>0$ be the privacy budget allocated to screening, let $\varepsilon_{num}>0$ be the privacy budget allocated to numerical measurements, and let $\delta\in(0,1)$ be the additive privacy parameter. The budgets are chosen such that the overall privacy parameter $\varepsilon$ satisfies

$$
\varepsilon_{sv}+\varepsilon_{num}\leq\varepsilon.
$$
Finally, let $C\in\mathbb{N}^*$ be the maximum number of allowed hypothesis updates.

We set the screening threshold to
$$
T=\frac{3\alpha}{4},
$$

leaving an $\alpha/4$ margin between the screening threshold and the target accuracy. This threshold is used by the screening procedure to determine whether a query requires a private numerical measurement. The choice $T=3\alpha/4$ is used throughout the proposed mechanism and its analysis.

\subsection{Sparse private screening}

The screening component follows the sparse-vector principle: rather than
obtaining a fresh private numerical measurement at every round, the
mechanism privately tests whether the current hypothesis is sufficiently
accurate and performs a numerical measurement only when the discrepancy
exceeds a prescribed threshold.

At round $t$, the screening statistic is
$$
g_t(D)=|q_t(D)-\mu_t|,
$$

which measures the discrepancy between the empirical answer $q_t(D)$ and
the current hypothesis answer $\mu_t$.

For each active update index $j\in\{0,\ldots,C-1\}$, draw threshold noise $\xi_j$ and, for each round, draw test noise $\nu_t$, according to the following Laplace distributions:
\begin{equation}
\xi_j \sim \operatorname{Lap}\left(\frac{2C}{n\varepsilon_{sv}}\right), \qquad
\nu_t \sim \operatorname{Lap}\left(\frac{4C}{n\varepsilon_{sv}}\right).
\end{equation}

Round $t$ is declared difficult if
$$
g_t(D)+\nu_t\geq T+\xi_j,
$$

where $j$ denotes the number of previous hypothesis updates. Otherwise, the
round is declared easy.

A difficult round triggers a fresh private numerical measurement and a subsequent update of the hypothesis. After each difficult round, $j$ is incremented and a new threshold noise $\xi_j$ is used. Once $j=C$, no further hypothesis updates are allowed.

\subsection{Private numerical measurement}

When a round is classified as difficult by the private screening step, the
mechanism obtains a fresh private numerical measurement of the empirical
query answer $a_t$. Since the query has sensitivity at most $1/n$ under
substitution adjacency, the mechanism adds Gaussian noise to protect the
privacy of this measurement. On the $j$-th difficult round, it releases
$$
y_t=a_t+Z_j,
\qquad
Z_j\sim\mathcal{N}\left(0,\frac{\sigma^2}{n^2}\right).
$$

The released value $y_t$ is therefore a noisy approximation of the empirical
answer $a_t$. The noise scale is calibrated according to the privacy budget
allocated to numerical measurements. Since at most $C$ difficult rounds can trigger a hypothesis update, we allocate the numerical-measurement privacy budget across these rounds by setting the per-measurement DP parameters as follows:
$$
\varepsilon_0=\frac{\varepsilon_{num}}{C},
\qquad
\delta_0=\frac{\delta}{C}.
$$

Assuming $0<\varepsilon_0<1$, we use the standard sufficient calibration for the Gaussian mechanism~\cite{dwork2014algorithmic} and choose
$$
\sigma =
\frac{1}{\varepsilon_0}
\sqrt{2\log\left(\frac{1.25}{\delta_0}\right)}
=
\frac{C}{\varepsilon_{num}}
\sqrt{2\log\left(\frac{1.25C}{\delta}\right)}.
$$

With this calibration, each numerical measurement satisfies an
$(\varepsilon_0,\delta_0)$-differential privacy guarantee for a query with
sensitivity $1/n$. By composition over at most $C$ numerical measurements,
the numerical-measurement component uses at most an $(\varepsilon_{num},\delta)$ privacy budget, and therefore satisfies $(\varepsilon_{num},\delta)$-DP.

\subsection{Variance-adaptive update}

On a difficult round, the mechanism uses the private numerical measurement
$y_t$ to determine how the current hypothesis distribution should be
corrected. In contrast to the true discrepancy $d_t=\mu_t-a_t$, which depends
on the private empirical answer, the update relies only on the privately
observed value $y_t$.

We therefore define the measured discrepancy $\widehat d_t$, its magnitude $\widehat e_t$, and its direction $\widehat s_t$, as follows:
$$
\widehat d_t=\mu_t-y_t, \qquad
\widehat e_t=|\widehat d_t|, \qquad
\widehat s_t=\operatorname{sign}(\widehat d_t).
$$

The magnitude of the update is controlled by an adaptive step size $\eta_t$ that depends on both the observed discrepancy $\widehat e_t$ and the local
variance $v_t$ of the query under the current hypothesis. Specifically,
$$
\eta_t= \begin{cases}
\log\left(1+\dfrac{\widehat e_t}{v_t}\right),
& v_t>0,\\[1.0ex] 0,
& v_t=0.
\end{cases}
$$

Thus, the step size adapts to the local variability of the query under
$h_t$, rather than using a fixed step size across all rounds.

The hypothesis distribution is then updated multiplicatively. For each
$x\in\mathcal{X}$, the updated distribution is given by
\begin{equation}
\label{eq:hypothesis-update}
h_{t+1}(x)=
\frac{h_t(x)\exp(-\eta_t\widehat s_t q_t(x))}
{\displaystyle\sum_{x'\in\mathcal{X}}h_t(x')
\exp(-\eta_t\widehat s_t q_t(x'))}.
\end{equation}

The updated distribution remains a probability distribution because the
denominator in \eqref{eq:hypothesis-update} normalizes the weights, ensuring that
$h_{t+1}\in\Delta(\mathcal{X})$. The sign $\widehat s_t$ determines the
direction of the correction, while the adaptive step size $\eta_t$
determines its magnitude.

Let $U\subseteq \{1,\ldots,k \}$ denote the set of rounds on which the hypothesis is updated, and let $m=|U|\leq C$ denote the total number of hypothesis updates. Thus, $m$ represents the number of difficult rounds that trigger a private numerical measurement and a subsequent hypothesis update, while $C$ is the maximum number of such updates allowed by the mechanism.

The entire mechanism is summarized in Algorithm~\ref{alg:pmw_bernstein}. A detailed privacy and potential analyses are presented in the next sections.

\begin{algorithm}[]
\caption{Proposed PMW-Bernstein framework}
\label{alg:pmw_bernstein}
\begin{algorithmic}[1]

\REQUIRE Private database $D\in\mathcal{X}^n$; adaptively chosen queries
$(q_t)_{t=1}^k$; prior $r\in\Delta(\mathcal{X})$; target accuracy $\alpha$;
DP parameters $\varepsilon,\delta$; privacy-budget allocation
$\varepsilon_{sv},\varepsilon_{num}$; update cutoff $C$.

\ENSURE Released transcript $(\widehat a_t)_{t=1}^{\ell}$ up to a stopping time $\ell\leq k$, and final hypothesis $h_{\ell+1}$.

\STATE Initialize $h_1\leftarrow r$, $j\leftarrow0$, and
$T\leftarrow3\alpha/4$.

\STATE Draw threshold noise
$\xi_0\sim\operatorname{Lap}(2C/(n\varepsilon_{sv}))$.

\FOR{$t=1,\ldots,k$}

\IF{$j=C$}
    \STATE \textbf{break}
\ENDIF

\STATE Compute $\mu_t\leftarrow\langle q_t,h_t\rangle$.

\STATE Set $v_t\leftarrow\operatorname{Var}_{h_t}(q_t)$.

\Comment{\textbf{Sparse private screening}}

\STATE Set $\text{easyRound}\leftarrow\TRUE$.

\IF{$j<C$}

    \STATE Draw
    $\nu_t\sim\operatorname{Lap}(4C/(n\varepsilon_{sv}))$.

    \STATE Set
    $g_t(D)\leftarrow|q_t(D)-\mu_t|$.

    \IF{$g_t(D)+\nu_t\geq T+\xi_j$}

        \STATE Set $\text{easyRound}\leftarrow\FALSE$.

    \ENDIF

\ENDIF

\IF{$\text{easyRound}=\TRUE$}

    \STATE Release $\widehat a_t\leftarrow\mu_t$.

    \STATE Set $h_{t+1}\leftarrow h_t$.

\ELSE

    \STATE 
    \Comment{\textbf{Private numerical measurement}}

    \STATE Set $j\leftarrow j+1$.

    \STATE Draw
    $Z_j\sim\mathcal{N}(0,\sigma^2/n^2)$.

    \STATE Set $y_t\leftarrow a_t+Z_j$.

    \Comment{\textbf{Variance-adaptive update}}

    \STATE Set
    $\widehat d_t\leftarrow\mu_t-y_t$,
    $\widehat e_t\leftarrow|\widehat d_t|$, and
    $\widehat s_t\leftarrow\operatorname{sign}(\widehat d_t)$.

    \IF{$v_t>0$}

        \STATE Set
        $\eta_t\leftarrow\log(1+\widehat e_t/v_t)$.

    \ELSE

        \STATE Set $\eta_t\leftarrow0$.

    \ENDIF

    \STATE Update $h_{t+1}$ according to Eq.~\eqref{eq:hypothesis-update}.

    \STATE Set $\widehat a_t\leftarrow y_t$.

    \IF{$j<C$}

        \STATE Draw
        $\xi_j\sim\operatorname{Lap}(2C/(n\varepsilon_{sv}))$.

    \ENDIF

\ENDIF

\ENDFOR
\end{algorithmic}
\end{algorithm}

\section{Privacy Analysis}
\label{sec:privacy-analysis}

In this section, we analyze the privacy guarantees of the proposed PMW-Bernstein framework. We first analyze the privacy of its main components, namely the sparse private screening, the private numerical measurement, and the variance-adaptive update. We then establish the overall privacy guarantee of the proposed framework by combining the privacy guarantees of the individual components under adaptive sequential composition. Finally, we show that, when the privacy budgets satisfy $\varepsilon_{sv}+\varepsilon_{num}\leq\varepsilon$, the proposed PMW-Bernstein framework satisfies $(\varepsilon,\delta)$-DP.

\begin{proposition}[Sparse-screening privacy]
\label{prop:sparse-screening-privacy}

The sparse private screening component described above satisfies $(\varepsilon_{sv},0)-DP$.
\end{proposition}

\begin{proof}
By Proposition~\ref{prop:sensitivity}, conditioned on the previously released transcript, the screening statistic is defined as
$$
g_t(D)=\left|q_t(D)-\mu_t\right|.
$$

For any neighboring databases $D\sim D'$, we have
$$
\left|g_t(D)-g_t(D')\right|\leq\frac{1}{n}.
$$

Thus, the sensitivity of the screening statistic is $\Delta_g=1/n$.

The screening procedure is an SVT with sensitivity $\Delta_g=1/n$ and cutoff $C$. Following the SVT privacy calibration presented in Definition~\ref{def:SVT}, the threshold and test noises are, respectively, sampled as
$$
\xi_j\sim\operatorname{Lap}\left(\frac{2C\Delta_g}{\varepsilon_{sv}}\right)
=
\operatorname{Lap}\left(\frac{2C}{n\varepsilon_{sv}}\right),
$$

and
$$
\nu_t\sim\operatorname{Lap}\left(\frac{4C\Delta_g}{\varepsilon_{sv}}\right)
=
\operatorname{Lap}\left(\frac{4C}{n\varepsilon_{sv}}\right).
$$

Therefore, by the privacy guarantee of the SVT construction, the screening transcript satisfies $(\varepsilon_{sv},0)$-DP.
\end{proof}

\begin{proposition}[Numerical-measurement privacy]
\label{prop:numerical-measurement-privacy}

Each difficult-round numerical release satisfies $(\epsilon_0,\delta_0)$-DP. Since at most $C$ such releases occur, their basic composition satisfies $(\epsilon_{num},\delta)$-DP.
\end{proposition}

\begin{proof}
By Proposition~\ref{prop:sensitivity}, conditioned on the previously released transcript, the empirical query answer $a_t=q_t(D)$ has sensitivity
$$
\Delta_q = \sup_{D\sim D'} \left|q_t(D)-q_t(D')\right| \leq\frac{1}{n}.
$$

As described in the private numerical measurement component, the mechanism adds Gaussian noise with standard deviation $\sigma/n$, where $\sigma$ is calibrated using the per-measurement DP parameters
$$
\varepsilon_0=\frac{\varepsilon_{num}}{C},
\qquad
\delta_0=\frac{\delta}{C}.
$$

Under the standard sufficient Gaussian-mechanism calibration given above, each numerical measurement therefore satisfies $(\varepsilon_0,\delta_0)$-DP.

Since at most $C$ difficult rounds can trigger a private numerical measurement, basic sequential composition gives
$$
(C\varepsilon_0,C\delta_0) = (\varepsilon_{num},\delta).
$$

Therefore, the numerical-measurement component uses at most an $(\varepsilon_{num},\delta)$ privacy budget, and therefore satisfies $(\varepsilon_{num},\delta)$-DP.
\end{proof}

\begin{proposition}[Privacy under adaptive query selection]
\label{prop:adaptive-query-privacy}

The privacy guarantees of the sparse private screening and private numerical-measurement components remain valid when the queries $(q_t)_{t=1}^k$ are adaptively chosen based on the previously released transcript.
\end{proposition}

\begin{proof}
At round $t$, the query $q_t$ may depend on the previously released transcript
$$
(\widehat a_1,\ldots,\widehat a_{t-1}).
$$

Conditioned on this transcript, the current query $q_t$ is fixed. By Proposition~\ref{prop:sensitivity}, its empirical answer $q_t(D)$ has sensitivity at most $1/n$. Consequently, the screening statistic
$$
g_t(D)=\left|q_t(D)-\mu_t\right|
$$

also has sensitivity at most $1/n$.

Therefore, conditioned on the previous transcript, the privacy guarantees established for the sparse private screening and private numerical-measurement components apply to the current round with the same privacy parameters. Since differential privacy is preserved under adaptive sequential composition, the adaptive selection of subsequent queries does not introduce any additional privacy loss.

Hence, the privacy guarantees of the proposed mechanism remain valid for the adaptively chosen query sequence $(q_t)_{t=1}^k$.
\end{proof}

\begin{theorem}[Overall privacy of PMW-Bernstein]
\label{thm:pmw-bernstein-overall-privacy}

The proposed PMW-Bernstein framework presented in Algorithm~\ref{alg:pmw_bernstein} is $(\varepsilon_{sv}+\varepsilon_{num},\delta)$-DP.
\end{theorem}

\begin{proof}

The privacy of the proposed PMW-Bernstein framework is based on two privacy-preserving components: the sparse private screening and the private numerical measurement.

By Proposition~\ref{prop:sensitivity} and Proposition~\ref{prop:sparse-screening-privacy}, the sparse private screening component satisfies $(\varepsilon_{sv},0)$-DP.  By Proposition~\ref{prop:numerical-measurement-privacy}, the private numerical measurement component satisfies $(\varepsilon_{num},\delta)$-DP. Proposition~\ref{prop:adaptive-query-privacy} ensures that these guarantees remain valid when the queries are adaptively selected based on the previously released transcript.

The two components (the sparse private screening and  the private numerical measurement) are applied sequentially and may depend on the previously released transcript. By the adaptive sequential composition property of differential privacy, their combined privacy loss is bounded by the sum of their respective privacy parameters. Hence,
$$
(\varepsilon_{sv},0)+(\varepsilon_{num},\delta)=
(\varepsilon_{sv}+\varepsilon_{num},\delta).
$$

The quantities used in the variance-adaptive update component (i.e., $\widehat d_t$, $\widehat e_t$, $\widehat s_t$, and $\eta_t$), as well as the updated hypothesis distribution $h_{t+1}$, are computed using the current query, the public hypothesis, and the privately released numerical measurement. Thus, the variance-adaptive update is a post-processing operation on the released information and incurs no additional privacy loss.

Therefore, the proposed PMW-Bernstein framework, as presented in Algorithm~\ref{alg:pmw_bernstein}, satisfies $(\varepsilon_{sv}+\varepsilon_{num},\delta)$-DP.
\end{proof}

\begin{corollary}[Overall differential privacy guarantee]
Suppose that the privacy-budget parameters satisfy
$$
\varepsilon_{sv}+\varepsilon_{num}\leq\varepsilon.
$$

Then the proposed PMW-Bernstein framework presented in Algorithm~\ref{alg:pmw_bernstein} satisfies $(\varepsilon,\delta)$-DP.
\end{corollary}

\begin{proof}
By Theorem~\ref{thm:pmw-bernstein-overall-privacy}, the proposed framework satisfies
$(\varepsilon_{sv}+\varepsilon_{num},\delta)$-DP. Since

$$
\varepsilon_{sv}+\varepsilon_{num}\leq\varepsilon,
$$
the privacy guarantee immediately implies $(\varepsilon,\delta)$-DP.
\end{proof}

\section{Potential Analysis}
\label{sec:potential-analysis}

This section analyzes the dynamics of the hypothesis update under the conditions that the noisy numerical release has the correct sign and provides a controlled multiplicative approximation to the true query error.

\subsection{Potential Function}

In this analysis, we use the classical PMW potential function, namely, the Kullback--Leibler (KL) divergence. The goal is to quantify the discrepancy between the hypothesis distribution $h_t(x)$ and the empirical distribution $p(x)$:
$$
\Phi_t=\KL(p\Vert h_t) =\sum_{x\in\calX}p(x)\log\frac{p(x)}{h_t(x)}.
$$

We assume that the initial hypothesis distribution $h_1$ has full support, i.e.,
$$
h_1(x)>0,\qquad \forall x\in\calX.
$$

The multiplicative update preserves this property, since every positive coordinate is multiplied by a positive exponential factor and then normalized. Consequently, $h_t$ has full support for all $t$, and the KL potential remains well-defined.

Recall that
$$
d_t=\mu_t-a_t, \qquad e_t=|d_t|, \qquad s_t=\sign(d_t),
$$

and
$$
v_t=\Var_{h_t}(q_t).
$$

If $v_t=0$, then $q_t$ is constant on $\calX$, and therefore
$$
a_t=\mu_t \qquad\text{and} \qquad e_t=0.
$$

Hence, for any updated round with $e_t>0$, we necessarily have $v_t>0$.

\subsection{Potential Decrease Under Accurate Measurement}

Consider an updated round and suppose that the noisy numerical release satisfies the correct-sign condition
$$
\widehat{s}_t=s_t=\sign(d_t),
$$

and provides a controlled multiplicative approximation to the true query error. Under the correct-sign condition, the hypothesis distribution update in Eq.~\eqref{eq:hypothesis-update} can be written as
$$
h_{t+1}(x) = \frac{h_t(x)\exp(-\eta_t s_t q_t(x))} {\mathbb{E}_{h_t}\!\left[\exp(-\eta_t s_t q_t)\right]}.
$$

The corresponding change in the KL potential is given by:
\begin{align}
\Phi_t-\Phi_{t+1} &= -\eta_t s_t a_t -\log\mathbb{E}_{h_t} \left[\exp(-\eta_t s_t q_t)\right].
\end{align}

Since $a_t=\mu_t-d_t$ and $s_t=\sign(d_t)$, we have
$$
-s_t a_t = -s_t\mu_t+e_t.
$$

Therefore,
\begin{align}
\Phi_t-\Phi_{t+1} &= \eta_t e_t -\eta_t s_t\mu_t -\log\mathbb{E}_{h_t} \left[\exp(-\eta_t s_t q_t)\right].
\end{align}

To express the exponential moment in terms of deviations from the hypothesis answer $\mu_t$, we write
$$
q_t=(q_t-\mu_t)+\mu_t.
$$

Thus,
\begin{equation}
\begin{split}
\mathbb{E}_{h_t}&\left[\exp(-\eta_t s_t q_t)\right] = \\
&\exp(-\eta_t s_t\mu_t) \mathbb{E}_{h_t}\left[\exp\left(-\eta_t s_t(q_t-\mu_t)\right)\right].
\end{split}
\end{equation}

Substituting this expression into the potential difference gives

\begin{align}
\Phi_t-\Phi_{t+1} &=
\eta_t e_t - \log\mathbb{E}_{h_t} \left[ \exp\left(-\eta_t s_t(q_t-\mu_t)\right)
\right].
\end{align}
Using the Bernstein-style exponential-moment bound for the centered and bounded query, we obtain
$$
\log\mathbb{E}_{h_t}
\left[
\exp\left(-\eta_t s_t(q_t-\mu_t)\right)
\right]
\leq
v_t\left(e^{\eta_t}-\eta_t-1\right).
$$
Consequently, the one-step decrease in the KL potential satisfies
\begin{equation}
\label{eq:fixed-progress}
\Phi_t-\Phi_{t+1} \geq \eta_t e_t - v_t\left(e^{\eta_t}-\eta_t-1\right).
\end{equation}
This inequality relates the decrease in the PMW potential to the true query error $e_t$ and the local query variance $v_t$. It provides the basis for analyzing the variance-adaptive step size
$$
\eta_t= \log\left(1+\frac{\widehat e_t}{v_t}\right),
$$
and for deriving a lower bound on the potential decrease under the assumed multiplicative approximation of the noisy query error.

\begin{lemma}[Logarithmic lower bound]
\label{lem:log-lower-bound}
For every $u\geq0$,
$$
\log(1+u)\geq \frac{2u}{2+u}.
$$
\end{lemma}

\begin{proof}
Let $f(u)=\log(1+u)-2u/(2+u)$. Since $f(0)=0$ and
$$
f'(u)=\frac{u^2}{(1+u)(2+u)^2}\geq0,
$$
the result follows.
\end{proof}

\begin{theorem}[Variance-adaptive progress]
\label{thm:adaptive_progress}
Consider an updated round with \(e_t>0\). Suppose the noisy measurement satisfies
\begin{equation}
\label{eq:favorable_measurement}
\widehat s_t=s_t,
\qquad
\widehat e_t\in\left[\frac{e_t}{2},\frac{3e_t}{2}\right].
\end{equation}

If the update uses \(\eta_t=\log(1+\widehat e_t/v_t)\), then
$$
\Phi_t-\Phi_{t+1}\geq \frac{3}{8}\frac{e_t^2}{v_t+e_t}.
$$
\end{theorem}

\begin{proof}
Let
$$
w=\frac{v_t}{e_t}>0, \qquad \rho=\frac{\widehat e_t}{e_t} \in \left[\frac12,\frac32\right].
$$
Then
$$
\eta_t=\log\left(1+\frac{\rho}{w}\right).
$$
Using Eq.~\eqref{eq:fixed-progress} and $e^{\eta_t}=1+\rho/w$ gives
\begin{align*}
\Phi_t-\Phi_{t+1}
&\geq e_t\left((1+w)\log\left(1+\frac{\rho}{w}\right)-\rho\right)\\
&=\frac{e_t}{w+1}F(w,\rho).
\end{align*}
where
$$
F(w,\rho)=(w+1)^2\log\left(1+\frac{\rho}{w}\right)-\rho(w+1).
$$
For fixed $w>0$,
$$
\frac{\partial F}{\partial\rho}
=\frac{(w+1)(1-\rho)}{w+\rho},
$$
so $F(w,\rho)$ increases on $(0,1]$ and decreases on $[1,\infty)$. Hence its minimum over $\rho\in[1/2,3/2]$ is attained at an endpoint.

We first compare the two endpoints. By Lemma~\ref{lem:log-lower-bound},
$$
\begin{aligned}
F\!\left(w,\frac32\right)-F\!\left(w,\frac12\right)
&=(w+1)^2\log\left(1+\frac{1}{w+1/2}\right)\\
&\quad -(w+1)\\
&\geq (w+1)^2\frac{1}{w+1}-(w+1)=0.
\end{aligned}
$$
Therefore the smaller endpoint is $\rho=1/2$. Applying Lemma~\ref{lem:log-lower-bound} again,
$$
\begin{aligned}
F\!\left(w,\frac12\right)
&=(w+1)^2\log\left(1+\frac{1}{2w}\right)-\frac12(w+1)\\
&\geq (w+1)^2\frac{2}{4w+1}-\frac12(w+1)\\
&=\frac{3(w+1)}{2(4w+1)}
=\frac38+\frac{9}{8(4w+1)}
\geq\frac38.
\end{aligned}
$$
Thus $F(w,\rho)\geq3/8$ for all $w>0$ and $\rho\in[1/2,3/2]$, and consequently
$$
\Phi_t-\Phi_{t+1}
\geq\frac38\frac{e_t}{w+1}
=\frac38\frac{e_t^2}{v_t+e_t}.
$$
\end{proof}

\subsection{Global Utility Bound}

The previous subsection established a lower bound on the decrease of the KL potential on an individual updated round. We now extend this one-step guarantee to all updated rounds and derive global bounds on the cumulative query error and on the number of updated rounds.

\begin{theorem}[Weighted energy bound]
\label{thm:energy}

Suppose \eqref{eq:favorable_measurement} holds for every updated round $t\in U$. Then
\begin{equation}
\sum_{t\in U}\frac{e_t^2}{v_t+e_t} \leq \frac{8}{3}\KL(p\Vert r).
\label{eq:energy}
\end{equation}
\end{theorem}

\begin{proof}
By Theorem~\ref{thm:adaptive_progress}, on every updated round $t\in U$,
$$
\Phi_t-\Phi_{t+1} \geq \frac{3}{8}\frac{e_t^2}{v_t+e_t}.
$$

Summing this inequality over all updated rounds gives
$$
\frac{3}{8}\sum_{t\in U}\frac{e_t^2}{v_t+e_t} \leq \sum_{t\in U}(\Phi_t-\Phi_{t+1}).
$$

Between two updated rounds, the hypothesis distribution remains unchanged. Therefore, the sum of the potential differences over the updated rounds telescopes to at most the total decrease of the potential from the initial distribution. Since the KL divergence is nonnegative,
$$
\sum_{t\in U}(\Phi_t-\Phi_{t+1}) \leq \Phi_1-\Phi_{\mathrm{final}} \leq \Phi_1.
$$

The initial hypothesis is $h_1=r$, and hence
$$
\Phi_1=\KL(p\Vert r).
$$

Combining the above inequalities yields
$$
\frac{3}{8} \sum_{t\in U}\frac{e_t^2}{v_t+e_t} \leq \KL(p\Vert r),$$

which gives
$$
\sum_{t\in U}\frac{e_t^2}{v_t+e_t} \leq \frac{8}{3}\KL(p\Vert r).
$$
\end{proof}

\begin{corollary}[Classical update regime]
\label{cor:classical}

Under the assumptions of Theorem~\ref{thm:energy},
$$
\sum_{t\in U}e_t^2 \leq \frac{10}{3}\KL(p\Vert r).
$$

If $e_t\geq\alpha$ on every updated round, then
$$
|U| \leq \frac{10}{3} \frac{\KL(p\Vert r)}{\alpha^2}.
$$
\end{corollary}

\begin{proof}
Since $q_t(x)\in[0,1]$, both $a_t$ and $\mu_t$ belong to $[0,1]$. Therefore,
$$
e_t=|a_t-\mu_t|\leq1.
$$

Moreover, since $q_t(x)\in[0,1]$, its variance under $h_t$ is bounded by the maximum variance of a random variable supported on $[0,1]$, namely
$$
v_t\leq\frac14.
$$

Consequently,
$$
v_t+e_t\leq\frac14+1=\frac54,
$$

and therefore
$$
\frac{e_t^2}{v_t+e_t} \geq \frac45e_t^2.
$$

Applying this inequality to the weighted energy bound in Theorem~\ref{thm:energy} gives
$$
\frac45\sum_{t\in U}e_t^2 \leq \frac83\KL(p\Vert r),
$$

which implies
$$
\sum_{t\in U}e_t^2 \leq \frac{10}{3}\KL(p\Vert r).
$$

If $e_t\geq\alpha$ for every $t\in U$, then 
$$
|U|\alpha^2 \leq \sum_{t\in U}e_t^2.
$$

Combining the two inequalities gives
$$
|U| \leq \frac{10}{3} \frac{\KL(p\Vert r)}{\alpha^2}.
$$
\end{proof}

\begin{corollary}[Bernstein-type fast regime]
\label{cor:bernstein}

Under the assumptions of Theorem~\ref{thm:energy}, suppose that for some $B\geq0$,
$$
v_t\leq B e_t \qquad \forall t\in U.
$$
Then
$$
\sum_{t\in U}e_t \leq \frac83(B+1)\KL(p\Vert r).
$$

If $e_t\geq\alpha$ on every updated round, then
$$
|U| \leq \frac83(B+1) \frac{\KL(p\Vert r)}{\alpha}.
$$
\end{corollary}

\begin{proof}
Under the assumed variance-error relation,
$$
v_t\leq Be_t,
$$

and hence
$$
v_t+e_t\leq(B+1)e_t.
$$

For $e_t>0$, it follows that
$$
\frac{e_t^2}{v_t+e_t} \geq \frac{e_t}{B+1}.
$$

Therefore, Theorem~\ref{thm:energy} gives
$$
\frac{1}{B+1} \sum_{t\in U}e_t \leq
\sum_{t\in U}\frac{e_t^2}{v_t+e_t} \leq
\frac83\KL(p\Vert r).
$$

Thus,
$$
\sum_{t\in U}e_t \leq \frac83(B+1)\KL(p\Vert r).
$$

Finally, if $e_t\geq\alpha$ for every $t\in U$, then
$$
|U|\alpha \leq \sum_{t\in U}e_t,
$$

which yields
$$
|U| \leq \frac83(B+1) \frac{\KL(p\Vert r)}{\alpha}.
$$
\end{proof}


\subsection{Screening-Calibrated Update Bounds}
\label{sec:screening-calibrated-bounds}

The update-complexity bounds derived in Corollaries~\ref{cor:classical} and~\ref{cor:bernstein} assume directly that every updated round satisfies $e_t\geq\alpha$. We now connect these conditional bounds to the noisy screening rule used by the complete mechanism. Consider the screening-calibration event
$$
|\xi_j|\leq\frac{\alpha}{8}
\quad\text{and}\quad
|\nu_t|\leq\frac{\alpha}{8}
$$
for all threshold and screening noises used by the mechanism.

Since $T=3\alpha/4$, any round declared difficult on this event satisfies
$$
e_t\geq T-|\xi_j|-|\nu_t| \geq \frac{\alpha}{2}.
$$

Thus, on the screening-calibration event, the conditional update bounds remain valid after replacing $\alpha$ by $\alpha/2$. Consequently, the corresponding mechanism-level update bounds are
$$
|U| \leq \frac{40}{3} \frac{\KL(p\Vert r)}{\alpha^2},
$$

and, under the condition $v_t\leq Be_t$ on updated rounds,
$$
|U| \leq \frac{16}{3}(B+1) \frac{\KL(p\Vert r)}{\alpha}.
$$

These bounds provide a direct connection between the theoretical update-complexity guarantees and the noisy screening procedure used by PMW-Bernstein. In particular, they quantify the effect of screening calibration on the constants in the classical and Bernstein-type update regimes.

\section{Derivation of PMW-Bernstein Key Components}
\label{sec:key-components-derivation}

The preceding analysis provides the theoretical basis for determining two key components of the PMW-Bernstein framework. The one-step potential bound identifies the step size that maximizes the guaranteed progress on an updated round, while the resulting global utility bounds determine a sufficient cutoff on the number of such updates. Accordingly, this section derives the variance-adaptive step and the update cutoff from the corresponding utility guarantees.

\subsection{Variance-Adaptive Step}
\label{sec:step-derivation}

The step used in the mechanism is obtained by maximizing the same one-step lower bound used in the potential proof when the error is known exactly. From \eqref{eq:fixed-progress}, the candidate-step objective is
$$
L_t(\eta) = \eta e_t-v_t(e^{\eta}-\eta-1).
$$

Differentiating with respect to $\eta$ gives
$$
L_t'(\eta) = e_t-v_t(e^{\eta}-1).
$$

Moreover,
$$
L_t''(\eta)=-v_t e^\eta<0,
$$

for $v_t>0$, so $L_t$ is strictly concave and its stationary point is the unique maximizer. Setting $L_t'(\eta)=0$ yields
$$
e^\eta=1+\frac{e_t}{v_t},
$$

and hence
$$
\eta_t^* = \log\left(1+\frac{e_t}{v_t}\right).
$$

The true error $e_t$ is inaccessible to the mechanism. We therefore replace it with the private estimate $\widehat e_t$, while using the privately determined sign $\widehat s_t$. The resulting adaptive step is thus based on the local ratio $\widehat e_t/v_t$.

Theorem~\ref{thm:adaptive_progress} shows that, despite the estimation error introduced by the private measurement, the resulting update retains the desired progress guarantee when $\widehat e_t$ approximates $e_t$ within a constant factor and the estimated sign is correct.

This also clarifies the relation with fixed-step PMW. A fixed step is selected independently of the local ratio $e_t/v_t$, requiring a compromise between stability and correction speed. In contrast, the variance-adaptive step responds to the local geometry of the update. That is, when $e_t\ll v_t$, we have
$$
\eta_t^* = \log\left(1+\frac{e_t}{v_t}\right) \approx \frac{e_t}{v_t},
$$

so the update is conservative. When $e_t\gg v_t$, the step grows only logarithmically,
$$
\eta_t^*
\sim
\log\left(\frac{e_t}{v_t}\right),
$$

rather than linearly in $e_t/v_t$. Thus, the adaptive rule automatically balances correction speed against the exponential cost appearing in the one-step potential bound.

\subsection{Update Cutoff Selection}

The cutoff $C$ has two roles. First, it bounds the number of numerical measurements, thereby determining the privacy allocation and the worst-case cumulative privacy cost of the numerical component. Second, it determines whether the mechanism can process the entire transcript before exhausting its update capacity.

The cutoff is a public mechanism parameter and therefore should not be chosen from the private quantity $\KL(p\Vert r)$ itself. Let $H_0$ be a public upper bound such that
$$
\KL(p\Vert r)\leq H_0.
$$
For example, when the prior is uniform on $\mathcal X$, one may take $H_0=\log|\mathcal X|$.

On the screening-calibration event discussed in Section~\ref{sec:screening-calibrated-bounds}, the general update bound gives the sufficient non-exhaustion scale
$$
C>\frac{40}{3}\frac{H_0}{\alpha^2}.
$$
Under the additional Bernstein-type condition $v_t\leq Be_t$ on updated rounds, the corresponding sufficient scale becomes
$$
C>\frac{16}{3}(B+1)\frac{H_0}{\alpha}.
$$
Thus, when a valid public upper bound on $B$ is available, the sufficient cutoff improves from order $1/\alpha^2$ to order $1/\alpha$. Without such prior information on $B$, the Bernstein result should be interpreted as a conditional statement about the update dynamics rather than as an automatic reduction of the privacy budget.

\section{Privacy--Utility Trade-off}
\label{sec:privacy-utility-tradeoff}

The privacy parameters directly affect the utility of the PMW-Bernstein mechanism through the noise introduced in the screening and numerical-measurement stages. In particular, stronger privacy requirements correspond to smaller privacy budgets and therefore larger perturbations. At the same time, the update cutoff $C$ plays an important role in this trade-off because it determines both the privacy allocation across numerical measurements and the magnitude of the noise used in the private measurement mechanism.

For the sparse private screening stage, the Laplace noise scales used are
$$
b_{\mathrm{thr}} = \frac{2C}{n\epsilon_{sv}},
\qquad
b_{\mathrm{test}} = \frac{4C}{n\epsilon_{sv}}.
$$

Thus, for fixed $n$ and $C$, decreasing $\epsilon_{sv}$ increases the screening noise and makes it more difficult to reliably distinguish accurate rounds from rounds requiring a private numerical measurement. Conversely, allocating a larger privacy budget to screening reduces the screening noise.

For the private numerical measurement stage, the Gaussian noise parameter and the resulting measurement noise variance are, respectively, given by
$$
\frac{C}{\epsilon_{num}} \sqrt{2\log\left(\frac{1.25C}{\delta}\right)}
\qquad\text{and}\qquad
\frac{\sigma^2}{n^2}.
$$

Hence, for fixed $C$, decreasing $\epsilon_{num}$ or $\delta$ strengthens the privacy guarantee but increases the perturbation of the released query answer. This increased noise can, in turn, affect the accuracy of the subsequent multiplicative update.

The cutoff $C$ creates an additional interaction between privacy and utility. Increasing $C$ allows the mechanism to accommodate more updated rounds, but it also increases the noise scales used in both the screening and numerical-measurement mechanisms. In particular, the numerical-measurement noise grows linearly with $C$, up to the logarithmic dependence on $C$ in the Gaussian calibration. Therefore, the cutoff should not be chosen unnecessarily large: it must be sufficiently large to accommodate the required number of updates while remaining consistent with the desired privacy--utility trade-off.

The screening-calibrated update bounds and the cutoff discussed above show that, on the screening-calibration event, the general sufficient cutoff has order $H_0/\alpha^2$, whereas the Bernstein-type regime has order $(B+1)H_0/\alpha$ when public upper bounds $H_0\geq\KL(p\Vert r)$ and $B$ are available. Since $C$ directly enters the privacy-noise calibration, a justified smaller cutoff can reduce the amount of noise introduced by the privacy mechanisms. This is a conditional consequence: without a public bound on $B$, the analysis does not by itself justify a smaller privacy allocation.

\section{Experimental Setup}
\label{sec:experimental-setup}

The preceding analysis characterizes the update dynamics and utility properties of the proposed PMW-Bernstein mechanism. To complement this theoretical analysis, we conduct controlled simulations to examine the behavior of the adaptive step under different variance and query-error regimes. The experiments compare the proposed adaptive PMW-Bernstein mechanism with a fixed-step PMW baseline under the same query-selection rules, target accuracy, and update budget.

\subsection{Hardware and Software}

The experiments were conducted on a computer running Microsoft Windows 11 Professional, 64-bit, version 10.0.26200. The system was equipped with an 11th Gen Intel(R) Core(TM) i7-1165G7 processor clocked at 2.80 GHz, with 4 physical cores and 8 logical processors, and approximately 16 GB of RAM. All experiments were conducted using Python 3.11.0.

\subsection{Datasets and Query Workloads}

The simulations are conducted using synthetic finite-universe distributions designed to evaluate the behavior of the proposed PMW-Bernstein mechanism under different variance and query-error regimes.

Each dataset is represented by a probability distribution $p$ over a finite universe $\mathcal{X}$ with $|\mathcal{X}|=128$. The initial hypothesis distribution $r$ is uniform over the universe, i.e.,
$$
r(x)=1/128, \qquad \forall x\in\mathcal{X}.
$$
Each query is represented by a vector in $[0,1]^{128}$, with its answer computed as the expectation of the corresponding query function under the true distribution. This representation follows directly from the mathematical formulation introduced in Section~\ref{sec:problem-statement} and allows the experiments to isolate the behavior of the adaptive update mechanism without relying on application-specific assumptions or real patient data. The simulation therefore operates directly on probability vectors; it does not generate a separate finite table of sampled records.

\subsection{Validation Scenarios}

The validation of the proposed mechanism is carried out through three scenarios, each representing a different variance regime. The parameters used to generate these scenarios are summarized in Table~\ref{tab:scenario-reproducibility}, which provides the necessary details for reproducibility.

\begin{table*}[]
\caption{Reproducibility parameters of the three synthetic scenarios used by the PMW comparison script.}
\label{tab:scenario-reproducibility}
\centering
\begin{tabularx}{\textwidth}{@{}p{2.9cm}Y Y Y@{}}
\toprule
Parameter & Low-variance regime & High-variance regime & Heterogeneous regime \\
\midrule
Universe and prior & $|\mathcal X|=128$; $r(x)=1/128$ & $|\mathcal X|=128$; $r(x)=1/128$ & $|\mathcal X|=128$; $r(x)=1/128$ \\
Synthetic empirical distribution
& Unnormalized weights: $14$ for elements $\{0,1,2,3\}$ and $1$ elsewhere; total weight $180$
& Unnormalized weights: $2.4$ for elements $\{0,\ldots,63\}$ and $0.6$ elsewhere; total weight $192$
& Gamma-generated weights: $w_i\overset{\mathrm{i.i.d.}}{\sim}\Gamma(0.35,1.0)$; normalized so that $p_i=w_i/\sum_j w_j$ \\

Scenario seed & $0$ & $1000$ & $2000$ \\
Query-pool size & $89$ & $98$ & $108$ \\
Query supports
& Four singleton indicators; structured supports of widths $2,4,8$ over the first 24 elements; 64 random sparse indicators with widths in $\{1,2,3,4,8\}$
& Two deterministic half-universe indicators and 96 random indicators of width $64$
& 18 random indicators at each density $0.02,0.04,0.08,0.16,0.32,0.5$, corresponding to widths $3,5,10,20,41,64$ \\

Variance
& Computed per round as $\Var_{h_t}(q_t)$; not a fixed dataset statistic
& Computed per round as $\Var_{h_t}(q_t)$; not a fixed dataset statistic
& Computed per round as $\Var_{h_t}(q_t)$; not a fixed dataset statistic \\
\bottomrule
\end{tabularx}
\end{table*}

The three scenarios, along with their expected outcomes, are presented below and summarized in Table~\ref{tab:experimental-datasets}.

\begin{enumerate}
\item \textbf{Scenario 1 (\emph{Low-variance regime}):} This scenario is designed to represent a regime in which the variance of the query under the current hypothesis is relatively small compared with the query error. The true distribution assigns elevated probability mass to four selected elements of $\mathcal X$ relative to the uniform prior ($r(x)$), while the query workload consists of single-element indicator queries and sparse random indicator queries. This configuration is intended to approximate the variance-error condition $v_t\leq B e_t$ considered in the Bernstein-type analysis and therefore provides a setting in which the adaptive step is expected to provide the greatest benefit.

\item \textbf{Scenario 2 (\emph{High-variance regime}):} This scenario represents a more balanced and higher-variance regime. The first half of the universe is assigned larger probability mass than the second half, and the query workload consists of half-universe indicator queries and random half-universe indicator queries. This scenario provides a contrasting setting in which the variance is relatively large compared with the query error, allowing the adaptive mechanism to be evaluated when the advantage predicted by the Bernstein-type analysis is less pronounced.

\item \textbf{Scenario 3 (\emph{Heterogeneous regime}):} This scenario uses a spiky true distribution generated from independent Gamma weights with shape $0.35$ and scale $1.0$, followed by normalization and combines sparse and dense random indicator queries at several densities. This heterogeneous workload produces varying relationships between the query error and variance across rounds and is therefore used to assess the robustness of the variance-adaptive step under more diverse query conditions.
\end{enumerate}

The proposed adaptive PMW-Bernstein mechanism is compared against a fixed-step PMW baseline whose step size is selected from the following grid:
$$
\eta \in \{0.25,0.5,1,2\}.
$$

The reported fixed-step baseline corresponds to the best-performing value in this grid, selected according to a hierarchical criterion: success rate is considered first, followed by the number of updates, and then the final maximum query error. This procedure provides a systematic comparison between the adaptive step and representative fixed-step choices without relying on an arbitrarily selected baseline parameter.

In all experiments, the target accuracy, number of query rounds, and maximum update count (cutoff) are, respectively, set to
$$
\alpha=0.03, \qquad k=120, \qquad C=60.
$$

The simulator uses an oracle maximum-error selection rule: at each round it selects the query in the fixed pool with the largest exact current error. This choice is intentionally non-private and is used only to isolate the update dynamics. A round is treated as difficult when the current exact error, plus the optional screening perturbation, is at least $\alpha$. The experimental simulator is therefore an ablation of the adaptive update rule rather than an end-to-end implementation of the Sparse Vector mechanism analyzed above.

With respect to perturbations, two experimental conditions are considered:
\begin{itemize}
\item \textbf{Noiseless condition:} Both screening and numerical-measurement perturbations are set to zero.
\item \textbf{Noisy condition:} The simulator adds Gaussian screening perturbation with standard deviation $0.01$ and Gaussian numerical-measurement perturbation with standard deviation $0.03$. These values define a robustness stress test and are not claimed to be the privacy-calibrated noise scales of Section~\ref{sec:proposed-solution}.
\end{itemize}
The three scenario distributions and query pools are generated once from the fixed scenario seeds reported in Table~\ref{tab:scenario-reproducibility}. For the noisy condition, 30 run seeds ($0$ through $29$) generate independent perturbation realizations. In the noiseless condition, the oracle selection rule and fixed scenarios make the 30 repetitions identical. The implementation caps the adaptive step at $\eta=20$ for numerical stability; this cap is inactive in the reported runs, for which the observed maximum adaptive step is below $2.5$. The corresponding implementation and tests are available in the public uchiDP repository.\footnote{\url{https://github.com/0xuchihamabben/uchiDP}}
\begin{table*}[]
\caption{Synthetic scenarios used in the experiments, along with the expected results.}
\label{tab:experimental-datasets}
\centering
\small
\begin{tabularx}{\textwidth}{@{}p{2cm}Y Y Y Y@{}}
\toprule
Scenario & True distribution & Query workload & Experimental purpose & Expected results \\
\midrule
Low-variance regime
&
Four selected elements of $\mathcal X$ have elevated probability mass relative to the uniform prior
&
Single-element indicator queries and sparse random indicator queries
&
Tests a low-variance, high-error regime
&
The adaptive step is expected to provide a greater advantage over fixed steps
\\
\midrule
High-variance regime
&
The first half of the universe has larger probability mass than the second half
&
Half-universe indicator queries and random half-universe indicator queries
&
Tests a higher-variance regime
&
The advantage of the adaptive step is expected to be less pronounced
\\
\midrule
Heterogeneous regime
&
A spiky distribution generated from a Gamma law
&
Sparse and dense random indicator queries at several densities
&
Tests a heterogeneous workload with varying variance-error relationships
&
The relative advantage of the adaptive step is expected to vary across query rounds
\\
\bottomrule
\end{tabularx}
\end{table*}

\section{Experimental Results}
\label{sec:experimental-results}

To evaluate the performance of the proposed PMW-Bernstein update rule, we consider three measures: the mean number of updates, the final maximum query-pool error, and the success rate over 30 runs for each scenario and perturbation setting.

For run $j$, let $U_j$ denote the set of update rounds, let $h_j^{\mathrm{final}}$ be the final hypothesis, and let $\mathcal Q$ be the fixed query pool of the scenario. The final maximum error is
$$
E_j^{\max}
=
\max_{q\in\mathcal Q}
\left|
\langle q,h_j^{\mathrm{final}}\rangle-\langle q,p\rangle
\right|.
$$
The reported means are
$$
\overline{U} = \frac{1}{30}\sum_{j=1}^{30}|U_j|, \qquad 
\overline{E}^{\max} = \frac{1}{30}\sum_{j=1}^{30}E_j^{\max}.
$$
The simulator enforces $|U_j|\leq C$ by construction, so the success rate used by the implementation is
$$
R_{\mathrm{succ}} =
\frac{1}{30}\sum_{j=1}^{30}
\mathbf{1}\!\left\{
E_j^{\max}\leq\alpha
\right\}.
$$
Thus, the experiment evaluates convergence of the final hypothesis on the entire fixed query pool; it does not measure the maximum error of all intermediate answers released during a private transcript.

Recall that we compare the proposed adaptive mechanism with a fixed-step PMW approach. Therefore, the \emph{Adaptive} configuration corresponds to the proposed PMW-Bernstein mechanism, in which the update step size is adjusted at each update round according to the current query error and variance. In contrast, the \emph{Fixed} configuration corresponds to the fixed-step PMW baseline, in which a single step size is selected from the grid $\{0.25,0.5,1,2\}$ and used throughout the updates. Thus, the comparison isolates the effect of the variance-adaptive step while keeping the remaining experimental settings unchanged. Under the stated hierarchical criterion, the selected fixed step is $\eta=0.5$ in the low- and high-variance scenarios and $\eta=0.25$ in the heterogeneous scenario. The obtained experimental results are summarized in Table~\ref{tab:pmw_results}.

\begin{table*}[]
\caption{Comparison of PMW-Bernstein and the best fixed-step PMW baseline across synthetic scenarios and perturbation settings. Reported values are averaged over 30 runs.}
\label{tab:pmw_results}
\centering
\begin{tabular}{l l r r r r r r}
\cmitoprule{3-8}
\multicolumn{2}{c}{} & \multicolumn{2}{c}{Mean updates} & \multicolumn{2}{c}{Mean Max. query error} & \multicolumn{2}{c}{Success rate} \\
\midrule
Scenario/Regime & Perturbation & Adaptive & Fixed & Adaptive & Fixed & Adaptive & Fixed \\
\hline
\multirow{2}{*}{Low-variance}
& Noiseless & 2.00 & 5.00 & 0.0048 & 0.0290 & 1.00 & 1.00 \\
& Noisy     & 2.30 & 5.00 & 0.0151 & 0.0290 & 1.00 & 1.00 \\
\hline
\multirow{2}{*}{High-variance}
& Noiseless & 2.00 & 3.00 & 0.0298 & 0.0176 & 1.00 & 1.00 \\
& Noisy     & 2.60 & 3.00 & 0.0157 & 0.0176 & 1.00 & 1.00 \\
\hline
\multirow{2}{*}{Heterogeneous}
& Noiseless & 28.00 & 60.00 & 0.0293 & 0.0319 & 1.00 & 0.00 \\
& Noisy     & 35.90 & 60.00 & 0.0286 & 0.0420 & 1.00 & 0.00 \\
\bottomrule
\end{tabular}
\end{table*}

We observe that in the low-variance regime, the adaptive rule reaches the target accuracy with two updates in the noiseless condition, compared with five updates for the best fixed-step baseline. Under the noisy condition, the adaptive rule also requires fewer updates on average while maintaining a lower maximum query error. These results are consistent with the variance-adaptive analysis, which predicts a greater benefit when the query variance is relatively small compared with the query error.

In the heterogeneous regime, the adaptive rule succeeds within the prescribed update budget in both perturbation settings, whereas the fixed-step baseline exhausts the budget and fails to reach the target accuracy. The adaptive mechanism also achieves a lower maximum query error in both settings. This result indicates that the adaptive step can remain effective when the relationship between query error and variance varies across query rounds.

In the high-variance regime, the two methods exhibit more similar update behavior and both achieve the target accuracy. This is consistent with the theoretical analysis: when $v_t$ is comparable to $e_t$, the variance-adaptive step approaches a fixed-step behavior, reducing the expected advantage of adaptivity.

\section{Discussion}
\label{sec:discussion}

The experimental results provide empirical evidence for the behavior predicted by the theoretical analysis of the proposed PMW-Bernstein mechanism. In particular, they illustrate how the relationship between the query error and the variance under the current hypothesis influences the effectiveness of the adaptive update rule.

In this section, we interpret these observations in relation to the theoretical guarantees, examine the Bernstein condition underlying the fast update regime, and consider the potential relevance of the mechanism to interactive privacy-preserving data analysis. We also discuss the broader theoretical significance of the variance-adaptive formulation and possible extensions of the privacy accounting.

\subsection{Interpretation of the Experimental Results}

The obtained experimental results are consistent with the theoretical analysis underlying the proposed PMW-Bernstein mechanism, particularly Theorem~\ref{thm:energy}. In the Low-variance regime, the current hypothesis assigns relatively low variance to the queried events, while the empirical distribution exhibits a visible discrepancy. This is the setting in which
$$
\eta_t=\log\left(1+\frac{\widehat e_t}{v_t}\right)
$$
can produce a relatively large correction because the estimated query error is large relative to the variance. This behavior is also reflected in the weighted progress term
$$
\frac{e_t^2}{v_t+e_t},
$$
which appears in the weighted progress bound of Theorem~\ref{thm:energy}.

In the High-variance regime, the adaptive and fixed-step baselines exhibit more similar behavior. In this setting, the variance is comparable to the query error, so the denominator $v_t+e_t$ is not dominated by the error term. Consequently, the advantage of adapting the step size to the local variance is less pronounced, which is consistent with the observed experimental results.

The Heterogeneous regime provides a more diverse setting in which the relationship between query error and variance changes across query rounds. The adaptive mechanism continues to achieve the target accuracy within the prescribed update budget, whereas the fixed-step baseline exhausts the budget without reaching the target. This result suggests that the adaptive rule can remain effective when the local conditions governing the update dynamics vary substantially across rounds.

\subsection{Bernstein Condition and Fast Update Regime}

The condition $v_t\leq B e_t$ stated in Corollary~\ref{cor:bernstein} is a local margin-type condition on the rounds that actually trigger updates. The variance $v_t$ measures dispersion under the current hypothesis, whereas $e_t$ measures the discrepancy between the current hypothesis and the empirical distribution. Thus, the condition identifies update rounds in which the query error is sufficiently large relative to the local variance.

A simple example in which the Bernstein condition can be satisfied is given by indicator queries. Let $q_t=\mathbf{1}_A$ denote the indicator query for a subset $A\subseteq\mathcal X$. Then
$$
v_t=\mu_t(1-\mu_t)\leq\mu_t.
$$

If the empirical value satisfies
$$
a_t\geq (1+\lambda)\mu_t
$$

for some $\lambda>0$, then
$$
e_t=a_t-\mu_t\geq\lambda\mu_t,
$$

and therefore
$$
v_t\leq\mu_t\leq\frac{1}{\lambda}e_t.
$$

In this case, the Bernstein condition holds with $B=1/\lambda$.

\subsection{Application Context}

The most natural applications are interactive statistical systems in which analysts ask many related subgroup queries over the same private population. Medical and demographic workloads often contain rare but important subpopulations. If the current hypothesis underestimates such a subgroup, the corresponding indicator query may have low variance under $h_t$ but a large empirical discrepancy. The proposed PMW-Bernstein mechanism is designed to exploit precisely this situation: it uses a private measurement of the query discrepancy, as in PMW, while adapting the correction magnitude to the local relationship between query error and variance.

\subsection{Theoretical Significance}

Rather than relying only on a uniform worst-case constant, the analysis establishes a weighted progress bound based on the term
$$
\frac{e_t^2}{v_t+e_t}.
$$

This yields a sharper update-complexity bound in rounds where the query error is large relative to the local variance. The improvement is therefore structural: it identifies the local conditions under which a stronger correction is justified by the geometry of the current hypothesis.

\subsection{Privacy Accounting Extensions}

The proposed mechanism was stated using simple Gaussian composition to keep the privacy analysis explicit. More advanced accounting techniques, such as zCDP or RDP, can be used for the numerical-measurement component to potentially reduce the resulting noise constants. The core potential argument remains unchanged because it depends on the accuracy and sign of the measurements used on update rounds.

\section{Conclusion}
\label{sec:conclusion}

This paper addressed the update inefficiency that can arise in the PMW approach when all detected errors are corrected with a fixed learning rate. The proposed PMW-Bernstein mechanism retains the standard privacy architecture of PMW: Sparse Vector screening determines when a private measurement is needed, Gaussian noise protects numerical measurements, and the hypothesis update is performed through post-processing. The new component is the variance-adaptive step, defined as
$$
\eta_t=\log(1+\widehat e_t/v_t).
$$

The theoretical analysis proves that, under the stated measurement condition, each updated round decreases the potential $\KL(p\Vert h_t)$ by at least
$$
\frac{3}{8}\frac{e_t^2}{v_t+e_t}.
$$

This yields a global weighted energy bound, recovers the classical $O(1/\alpha^2)$ update regime in general, and establishes an $O(1/\alpha)$ update bound under a Bernstein-type condition.

Controlled simulations support the theoretical interpretation. The adaptive rule reduces the number of updates in the low-variance regime, remains competitive in the high-variance regime, and achieves the target accuracy within the prescribed update budget in the heterogeneous regime where the fixed-step baseline fails. Overall, the proposed mechanism provides a variance-adaptive refinement of PMW's update dynamics while retaining the differential privacy framework of the underlying mechanism.



\begin{thebibliography}{00}
\label{sec:bibliography}
\balance

\bibitem{dwork2006calibrating}
C. Dwork, F. McSherry, K. Nissim, and A. Smith, ``Calibrating noise to sensitivity in private data analysis,'' in \emph{Proc. Theory of Cryptography Conference}, 2006, pp. 265--284.

\bibitem{dwork2014algorithmic}
C. Dwork and A. Roth, ``The algorithmic foundations of differential privacy,'' \emph{Foundations and Trends in Theoretical Computer Science}, vol. 9, no. 3--4, pp. 211--407, 2014.

\bibitem{dwork2010boosting}
C. Dwork, G. N. Rothblum, and S. Vadhan, ``Boosting and differential privacy,'' in \emph{Proc. IEEE Symposium on Foundations of Computer Science}, 2010, pp. 51--60.

\bibitem{bun2016concentrated}
M. Bun and T. Steinke, ``Concentrated differential privacy: Simplifications, extensions, and lower bounds,'' in \emph{Proc. Theory of Cryptography Conference}, 2016, pp. 635--658.

\bibitem{mironov2017renyi}
I. Mironov, ``R\'enyi differential privacy,'' in \emph{Proc. IEEE Computer Security Foundations Symposium}, 2017, pp. 263--275.

\bibitem{lyu2017understanding}
M. Lyu, D. Su, and N. Li, ``Understanding the sparse vector technique for differential privacy,'' \emph{Proceedings of the VLDB Endowment}, vol. 10, no. 6, pp. 637--648, 2017.

\bibitem{blum2008learning}
A. Blum, K. Ligett, and A. Roth, ``A learning theory approach to noninteractive database privacy,'' in \emph{Proc. ACM Symposium on Theory of Computing}, 2008, pp. 609--618.

\bibitem{hardt2010multiplicative}
M. Hardt and G. N. Rothblum, ``A multiplicative weights mechanism for privacy-preserving data analysis,'' in \emph{Proc. IEEE Symposium on Foundations of Computer Science}, 2010, pp. 61--70.

\bibitem{roth2010interactive}
A. Roth and T. Roughgarden, ``Interactive privacy via the median mechanism,'' in \emph{Proc. ACM Symposium on Theory of Computing}, 2010, pp. 765--774.

\bibitem{hardt2012simple}
M. Hardt, K. Ligett, and F. McSherry, ``A simple and practical algorithm for differentially private data release,'' in \emph{Advances in Neural Information Processing Systems}, 2012, pp. 2348--2356.

\bibitem{mcsherry2009privacy}
F. McSherry, ``Privacy integrated queries: An extensible platform for privacy-preserving data analysis,'' in \emph{Proc. ACM SIGMOD International Conference on Management of Data}, 2009, pp. 19--30.

\bibitem{dwork2009complexity}
C. Dwork, M. Naor, O. Reingold, G. N. Rothblum, and S. Vadhan, ``On the complexity of differentially private data release: Efficient algorithms and hardness results,'' in \emph{Proc. ACM Symposium on Theory of Computing}, 2009, pp. 381--390.


\bibitem{mcsherry2007mechanism}
F. McSherry and K. Talwar, ``Mechanism design via differential privacy,'' in \emph{Proc. IEEE Symposium on Foundations of Computer Science}, 2007, pp. 94--103.

\bibitem{li2010matrix}
C. Li, M. Hay, V. Rastogi, G. Miklau, and A. McGregor, ``Optimizing linear counting queries under differential privacy,'' in \emph{Proc. ACM Symposium on Principles of Database Systems}, 2010, pp. 123--134.

\bibitem{barak2007privacy}
B. Barak, K. Chaudhuri, C. Dwork, S. Kale, F. McSherry, and K. Talwar, ``Privacy, accuracy, and consistency too: A holistic solution to contingency table release,'' in \emph{Proc. ACM Symposium on Principles of Database Systems}, 2007, pp. 273--282.

\bibitem{liu2021iterative}
T. Liu, G. Vietri, and S. Z. Wu, ``Iterative methods for private synthetic data: Unifying framework and new methods,'' in \emph{Advances in Neural Information Processing Systems}, vol. 34, 2021.

\bibitem{mckenna2022aim}
R. McKenna, B. Mullins, D. Sheldon, and G. Miklau, ``AIM: An adaptive and iterative mechanism for differentially private synthetic data,'' \emph{Proc. VLDB Endowment}, vol. 15, no. 11, pp. 2599--2612, 2022.

\end{thebibliography}
\end{document}